\documentclass[aps,pra,floatfix,amsmath,amssymb,superscriptaddress,showpacs,showkeys,twocolumn,10pt]{revtex4-1}
\usepackage[caption=false]{subfig}
\usepackage{graphicx,bm,color}
\usepackage{amsfonts}
\usepackage{color}
\usepackage{algorithm}
\usepackage{algpseudocode}
\usepackage{braket}
\usepackage[dvipsnames]{xcolor}
\usepackage{hyperref}
\hypersetup{
    colorlinks=true,
    linkcolor=blue,
    filecolor=blue,      
    urlcolor=blue,
}
\usepackage{epstopdf}

\begin{document}

\frenchspacing

\title{Hyper-hybrid entanglement, indistinguishability, and two-particle entanglement swapping}
\author{Soumya Das}
\email{soumya06.das@gmail.com}
\affiliation{Cryptology and Security Research Unit, R. C. Bose Centre for Cryptology and Security, Indian Statistical Institute, Kolkata 700108, India}
\author{Goutam Paul}
\email{goutam.paul@isical.ac.in}
\affiliation{Cryptology and Security Research Unit, R. C. Bose Centre for Cryptology and Security, Indian Statistical Institute, Kolkata 700108, India}
\author{Anindya Banerji}
\email{abanerji09@gmail.com}
\affiliation{Quantum Science and Technology Laboratory, Physical Research Laboratory, Ahmedabad 380009, India}

\begin{abstract}
Hyper-hybrid entanglement for two indistinguishable bosons has been recently proposed by Li \textit{et al.} [Y. Li, M. Gessner, W. Li, and A. Smerzi, \href{https://doi.org/10.1103/PhysRevLett.120.050404}{Phys. Rev. Lett. 120, 050404 (2018)}]. In the current paper, we show that this entanglement exists for two indistinguishable fermions also. Next, we establish two {\em no-go} results: no hyper-hybrid entanglement for two {\em distinguishable} particles, and no unit fidelity quantum teleportation using {\em indistinguishable} particles. If either of these is possible, then the {\em no-signaling principle} would be violated. While several earlier works have attempted extending many results on distinguishable particles to indistinguishable ones, and vice versa, the above two  no-go results establish a nontrivial separation between the two domains. Finally, we propose an efficient entanglement swapping using only two indistinguishable particles, whereas a minimum number of either three distinguishable or four indistinguishable particles is necessary for existing protocols.
\end{abstract}

\maketitle

\section{Introduction} 
In the last century, physicists were puzzled about whether ``the characteristic trait of Quantum Mechanics"~\cite{Schrodinger}, i.e., entanglement~\cite{epr}, is real and, if so, whether it can show some nontrivial advantages over classical information processing tasks. The answers to both are positive, thanks to several experimentally verified quantum protocols like teleportation~\cite{QT93,QTnat15}, dense coding~\cite{DC92,Mattle96}, quantum cryptography,~\cite{BB84,QKDexp92} etc.~\cite{HHHH08}.

In the current century, entanglement of indistinguishable particles and its similarity with as well as difference from that of distinguishable ones have been extensively studied~\cite{Li01,You01,John01,Zanardi02,Ghirardhi02,Wiseman03,Ghirardi04,Vedral03,Barnum04,Barnum05,Zanardi04,Omar05,Eckert02,Grabowski11,Sasaki11,Tichy13,Kiloran14,Benatti17,LFC16,Braun18,LFC18}. Here, indistinguishable particles means independently prepared identical particles like bosons or fermions~\cite{Feynman94,Sakurai94}, where each particle cannot be addressed individually, i.e., a label cannot be assigned to each.
Experiments on quantum dots~\cite{Petta05,Tan15}, Bose-Einstein condensates~\cite{Morsch06,Esteve08}, ultracold atomic gases~\cite{Leibfried03}, etc., support the existence of entanglement of indistinguishable particles. 

The notion of entanglement for distinguishable particles is well studied in the literature~\cite{HHHH08}, where the standard bipartite entanglement is  measured by Schmidt coefficients~\cite{Nielsenbook}, von Neumann entropy~\cite{Bennett96}, concurrence~\cite{CKW00}, log negativity~\cite{Vidal02}, etc.~\cite{Plenio07}. 
Indistinguishability, on the other hand, is represented and analyzed via particle-based first quantization approach~\cite{Li01,You01,John01,Zanardi02,Ghirardhi02,Wiseman03,Ghirardi04} or mode-based second-quantization approach~\cite{Vedral03,Barnum04,Barnum05,Zanardi04}. 
Entanglement in such a scenario requires measures~\cite{Eckert02,Grabowski11,Sasaki11,Tichy13,Kiloran14,Benatti17,LFC16} different from those of distinguishable particles, but there is no consensus on this in the scientific community~\cite[Sec.~III]{Braun18}, particularly on the issues of physicality~\cite{Zanardi02,Barnum04}, accessibility~\cite{Esteve08,Eckert02}, and usefulness~\cite{Tichy13,Kiloran14} of such entanglement. Very recently, the resource theory of indistinguishable particles~\cite{LFC16,LFC18} has been proposed aiming to settle this debate.

In order to treat the entanglement of distinguishable as well as indistinguishable particles on an equal footing, we use the algebraic framework introduced in~\cite{Benatti10,Benatti12,Benatti14} (see Appendix  for more details). To measure this entanglement in the case of both distinguishable and indistinguishable particles, we use the violation of Clauser-Horne-Shimony-Holt (CHSH)~\cite{CHSH} inequalities.

\begin{figure}[t!]
\centering
\includegraphics[width=8.6cm]{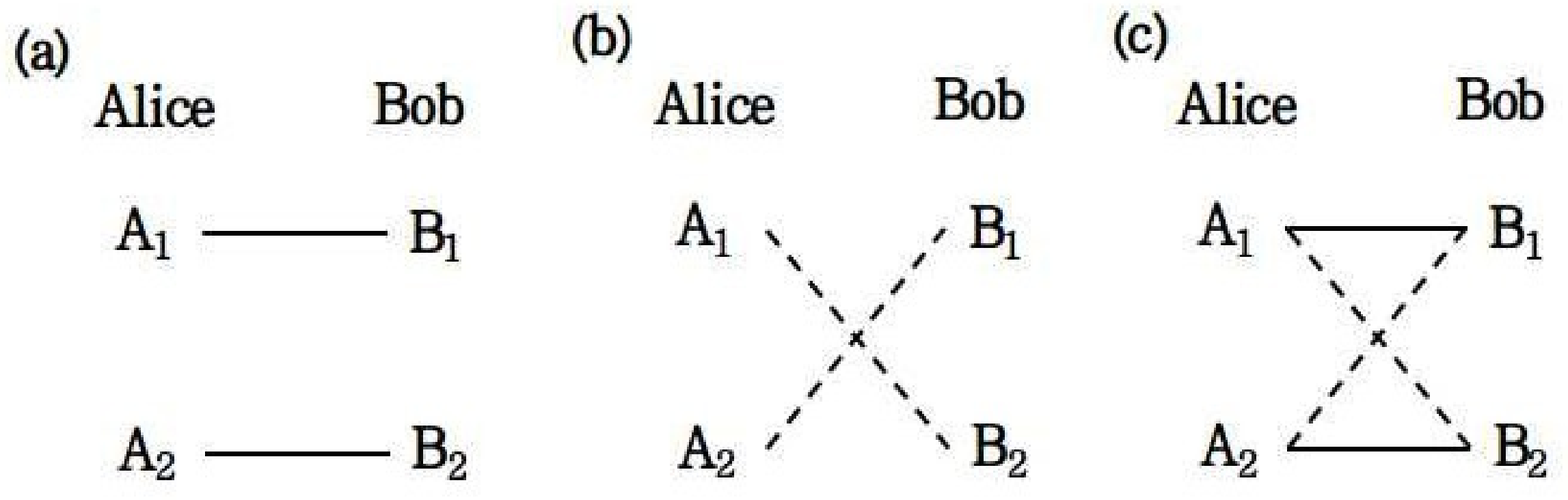} 
\caption{From left to right: (a) hyper-entanglement (solid lines), (b) hybrid-entanglement (dotted lines), and (c) hyper-hybrid entangled state of two qubits with two degrees of freedom.}
\label{fig:HHES}
\end{figure}

Quantum entanglement is encoded in the particle's degrees of freedom (DOFs)  like spin, polarization, path, angular momentum, etc. The simultaneous presence of entanglement in multiple DOFs, i.e., hyper-entanglement~\cite{Kwait97}, is useful for some tasks like complete Bell state analysis~\cite{Sheng10}, entanglement concentration ~\cite{Ren13}, purification~\cite{Ren13b}, etc.~\cite{Deng17}. Also, entanglement between different DOFs, i.e., hybrid entanglement~\cite{Zukowski91,Ma09}, is useful in quantum repeaters~\cite{Loock06}, quantum erasers~\cite{Ma13}, quantum cryptography~\cite{Sun11}, etc.~\cite{chen02,Neves09,Andersen15}. Although hyper-entanglement and hybrid entanglement were separately known for more than two decades, interestingly, the simultaneous presence of these two, i.e., the hyper-hybrid entangled state (HHES), has been proposed very recently by Li \textit{et al.}~\cite{HHNL}, using particle exchange~\cite{Y&S92PRA,Y&S92PRL} method. The above three types of entanglements are shown schematically in Fig.~\ref{fig:HHES}.

This discussion raises a few natural questions. The first two questions are as follows. 

 {\em (i) Is HHES possible for two indistinguishable fermions also}? 

 {\em (ii) Is the scheme for HHES, as proposed by  Li \textit{et al.}, applicable for two distinguishable particles}? 

\noindent Systematic calculations establish the answer to the first question as positive and that to the second as negative. The negative answer (i.e., the specific scheme of Li \textit{et al.}~\cite{HHNL} does not have a distinguishable version) immediately leads to the obvious but nontrivial {\em third} question. 

 {\em (iii) Can distinguishable particles exhibit HHES through some other scheme}? 

\noindent In answer, we establish the following \textit{no-go} result: HHES is not possible for distinguishable particles;	 otherwise, exploiting it, signaling can be achieved. 
Throughout this paper, by signaling we mean faster-than-light or superluminal communication across spacelike separated regions.	

 Bose and Home~\cite{Bose13} have identified a property of indistinguishable particles, called the {\em duality of entanglement}, which, according to them, is absent in distinguishable particles. However, Karczewski and Kurzy\'{n}ski~\cite{M.Karczewski16} have shown later that the duality of entanglement can be seen in distinguishable particles also. Thus, so far HHES remains the exclusive property that is possible for indistinguishable particles only.

As HHES is unique to indistinguishable particles, the following {\em fourth} question naturally comes up. 

 {\em (iv) Is there something unique in the case of distinguishable particles}?

\noindent The answer is positive, due to our second no-go result: unit fidelity quantum teleportation (UFQT)~\cite{Popescu94} is not possible for indistinguishable particles; otherwise, using it, the no-signaling principle can be violated.

The above two no-go results establish a separation between quantum properties and applications of distinguishable particles and those of indistinguishable ones.

Apart from the above unique properties and applications, there are many, such as coherence~\cite{Baumgratz14,Sperling17}, entanglement swapping (ES)~\cite{ES93,LFCES19}, metrology~\cite{Cronin09,Benatti10}, steering~\cite{Schrodinger,Fadel18}, etc., that are common to both distinguishable and indistinguishable particles. The seminal work~\cite{ES93} on ES required four distinguishable particles as a resource along with Bell state measurement (BSM)~\cite{BSM99} and local operations and classical communications (LOCC)~\cite{LOCC} as tools. Better versions with only three distinguishable particles were proposed in two subsequent works, one~\cite{Pan10} with BSM and another~\cite{Pan19} without BSM. Recently, Castellini \textit{et al.}~\cite{LFCES19} have shown that ES for the indistinguishable case is also possible with four particles (with BSM for bosons and without BSM for fermions). 

Thus, in terms of resource requirement, the existing best distinguishable versions~\cite{Pan10,Pan19} outperform the indistinguishable one~\cite{LFCES19}. We turn around this view, by proposing an ES protocol without BSM using only two indistinguishable particles.

This paper is arranged as follows. In Sec.~\ref{HHESfermions}, we show the existence of HHES for two indistinguishable fermions and also present a unified mathematical formalism for HHES that includes both bosons and fermions. Sec.~\ref{DisHHES} shows the inapplicability of the circuit of Li \textit{et al.}~\cite{HHNL} for two distinguishable particles. We present our first no-go result, i.e., no HHES for distinguishable particles, in Sec.~\ref{NoHHES}. Here we also propose a generic signaling scheme using HHES for distinguishable particles, which is an application of the above no-go result. However, the above generic scheme requires a large number of DOFs and thus may be hard to realize experimentally. As a remedy, we show an experimentally realizable scheme using only two DOFs. Sec.~\ref{NoUFQT} discusses our second no-go result, i.e., no unit fidelity quantum teleportation for indistinguishable particles. Using the above two no-go results, we illustrate nontrivial separation between the quantum properties and applications of distinguishable and indistinguishable particles in Sec.~\ref{Sepresult}. In Sec.~\ref{2PES}, we propose a circuit to perform the entanglement swapping protocol using only two particles and without using BSM. Finally, in Sec~\ref{Conclusion}, we discuss and summarize all the results and their implications.

\section{Hyper-Hybrid Entangled State for two indistinguishable fermions} \label{HHESfermions} 
Yurke and Stolar~\cite{Y&S92PRA,Y&S92PRL} had proposed an optical circuit to generate quantum entanglement between the same DOFs of two identical particles (bosons and fermions) from initially separated independent sources. 
Recently, the above method has been extended by Li \textit{et al.}~\cite{HHNL} to generate HHES between two independent bosons among their internal (e.g., spin) DOFs, external (e.g., momentum) DOFs, and across. We show that their circuit can also be used for independent fermions obeying the Pauli exclusion principle~\cite{Pauli25}, albeit with different detection probabilities.

For fermions, the second quantization formulation deals with fermionic creation operators $f_{i,\textbf{p}}$ with $\ket{i,\textbf{p}}=f^{\dagger}_{i,\textbf{p}}\ket{0}$, where $\ket{0}$ is the vacuum and $\ket{i,\textbf{p}}$ describes a particle with spin $\ket{i}$ and momentum $\textbf{p}$. These operators satisfy the canonical anticommutation relations:
\begin{equation} 
\left\lbrace f_{i,\textbf{p}_{i}}, f_{j,\textbf{p}_{j}}\right\rbrace = 0, \hspace{0.2cm} \left\lbrace f_{i,\textbf{p}_{i}}, f^{\dagger}_{j,\textbf{p}_{j}}\right\rbrace = \delta\left( \textbf{p}_{i}-\textbf{p}_{j}\right)\delta_{ij}.
\end{equation}

Analysis of the circuit of Li \textit{et al.}~\cite[Fig. 2]{HHNL} for fermions involves an array of hybrid beam splitters (HBSs)~\cite[Fig. 3]{HHNL}; phase shifter; four orthogonal external modes $L$, $D$, $R$, and $U$; and two orthogonal internal modes $\uparrow$ and $\downarrow$. Here, particles exiting through the modes $L$ and $D$ are received by Alice (A), who can control the phases $\phi_{L}$ and $\phi_{D}$, whereas particles exiting through the modes $R$ and $U$  are received by Bob (B), who can control the phases $\phi_{R}$ and $\phi_{U}$. 

In this circuit~\cite[Fig. 2]{HHNL}, two particles, each with spin $\ket{\downarrow}$, enter the setup in the mode $R$ and $L$ for Alice and Bob, respectively. The initial state of the two particles is $\ket{\Psi_{0}}=f^{\dagger}_{\downarrow,R} f^{\dagger}_{\downarrow,L}\ket{0}$. Now, the particles are sent to the HBS such that one output port of the HBS is sent to the other party ($R$ or $L$) and the other port remains locally accessible ($D$ or $U$). Next, each party applies path-dependent phase shifts. Lastly, the output of the local mode and that received from the other party are mixed with the HBS and then the measurement is performed in either  external or internal modes. The final state can be written as
\begin{equation} \label{Final_state_fermion}
\begin{aligned}
&\ket{\Psi}=\frac{1}{4} \left[ e^{i\phi_{R}}\left( f^{\dagger}_{\downarrow,R}+if^{\dagger}_{\uparrow,U}\right) +ie^{i\phi_{D}} \left( f^{\dagger}_{\uparrow,D} + i f^{\dagger}_{\downarrow,L}\right)  \right] \\
& \otimes \left[ e^{i\phi_{L}} \left( f^{\dagger}_{\downarrow,L}+if^{\dagger}_{\uparrow,D}\right)  + i e^{i\phi_{U}}\left( f^{\dagger}_{\uparrow,U}+if^{\dagger}_{\downarrow,R}\right) \right] \ket{0}.
\end{aligned}
\end{equation} 
Alice and Bob can perform coincidence measurements both in external DOFs or both in internal DOFs or with one party in the internal DOF and the other in the external DOF.  Now from Eq.~\eqref{Final_state_fermion}, the detection probabilities when each party gets exactly one particle where both Alice and Bob measure in external DOFs are given by 
\begin{equation}\label{path_path}
  \begin{tabular}{c | c c} 
  & $B$ : $R$ & $B$ : $U$  \\  
 \hline
 $A$ : $D$ \hspace{0.0005cm} & \hspace{0.0005cm} $\frac{1}{4}\text{cos}^2\phi$  & $\frac{1}{4}\text{sin}^2\phi$ \\ 
 $A$ : $L$ \hspace{0.0005cm} & \hspace{0.0005cm} $\frac{1}{4}\text{sin}^2\phi$  & $\frac{1}{4}\text{cos}^2\phi$ \\
\end{tabular},
\end{equation}
where $\phi=\left( \phi_{D} -\phi_{L} - \phi_{R}+ \phi_{U}\right) / 2$.

Now we assign dichotomic variables $+1$ and $-1$ for the detection events $\lbrace L,U \rbrace$ and $\lbrace D,R \rbrace$, respectively. Let $Pr_{mn}$ denote the probabilities of the coincidence events for Alice and Bob obtaining $m=\pm 1$ and $n=\pm 1$, respectively. The normalized expectation value is then given by
\begin{align} \label{Normal}
    E\left( \phi_{A},\phi_{B} \right)= &\dfrac{Pr_{++}-Pr_{-+}-Pr_{+-}+Pr_{--}}{Pr_{++}+Pr_{-+}+Pr_{+-}+Pr_{--}} \nonumber \\
    =& \text{cos} \left( \phi_{A} -\phi_{B} \right),
\end{align}
\noindent where $\phi_{A}=\left(\phi_{D}-\phi_{L}\right) $ and $\phi_{B}=\left( \phi_{U}-\phi_{R} \right)$. Now the CHSH~\cite{CHSH} inequality can be written as 
\begin{equation} \label{CHSH}
\vert E\left( \phi^{0}_{A},\phi^{0}_{B} \right) + E\left( \phi^{1}_{A},\phi^{0}_{B} \right)+ E\left( \phi^{0}_{A},\phi^{1}_{B} \right) - E\left( \phi^{1}_{A},\phi^{1}_{B} \right) \vert \leq 2,
\end{equation}
where the superscripts $0$ and $1$ stand for two detector settings for each particles. Now for $\phi^{0}_{A}=0$, $\phi^{1}_{A}=\pi$, $\phi^{0}_{B}=\frac{\pi}{4}$, and $\phi^{1}_{B}=-\frac{\pi}{4}$, Eq.~\eqref{CHSH} can be violated maximally by obtaining Tsirelson's bound $2\sqrt{2}$~\cite{Tsirelson}.

Now if Alice and Bob both measure	 in internal DOFs, then  the detection probabilities can be written as
\begin{equation}\label{spin_spin}
   \begin{tabular}{c | c c} 
  & $B$ : $\downarrow$ & $B$ : $\uparrow$  \\  
 \hline\
 $A$ : $\downarrow$  & \hspace{0.0005cm}  $\frac{1}{4}\text{sin}^2\phi$ \hspace{0.1cm} & $\frac{1}{4}\text{cos}^2\phi$\\ 
 \hspace{0.001cm} $A$ : $\uparrow$  & \hspace{0.0005cm}  $\frac{1}{4}\text{cos}^2\phi$ \hspace{0.1cm} & $\frac{1}{4}\text{sin}^2\phi$ \\
\end{tabular}.
\end{equation}
If Alice measures in the internal DOF and Bob measures in  the external DOF, then  the detection probabilities can be written as 
\begin{equation} \label{spin_path}
\begin{tabular}{c | c c} 
  & $B$ : $R$ & $B$ : $U$  \\  
 \hline
 $A$ : $\downarrow$  & \hspace{0.0005cm} $\frac{1}{4}\text{sin}^2\phi$ \hspace{0.1cm} & $\frac{1}{4}\text{cos}^2\phi$\\ 
 $A$ : $\uparrow$ & \hspace{0.0005cm} $\frac{1}{4}\text{cos}^2\phi$ \hspace{0.1cm} & $\frac{1}{4}\text{sin}^2\phi$ \\
\end{tabular}.
\end{equation}
If Alice measures in the external DOF and Bob measures in the internal DOF, then the detection probabilities can be written as 
\begin{equation} \label{path_spin}
\begin{tabular}{c | c c} 
  & $B$ : $\downarrow$ & $B$ : $\uparrow$  \\  
 \hline\
 $A$ : $D$  & \hspace{0.0005cm} $\frac{1}{4}\text{cos}^2\phi$ \hspace{0.1cm} & $\frac{1}{4}\text{sin}^2\phi$\\ 
 $A$ : $L$  & \hspace{0.0005cm} $\frac{1}{4}\text{sin}^2\phi$  \hspace{0.1cm} & $\frac{1}{4}\text{cos}^2\phi$ \\
\end{tabular}.
\end{equation}
 Now by applying similar analysis for Eqs.~\eqref{spin_spin},~\eqref{spin_path}, and~\eqref{path_spin}  as performed for Eqs.~\eqref{Normal} and~\eqref{CHSH}, one can show maximal violation of Bell's inequality. 

\subsection{Generalized hyper-hybrid entangled state}
Interestingly, following the approach by Yurke and Stolar~\cite{Y&S92PRA},
we can generalize the detection probabilities of HHES for indistinguishable bosons and fermions into a single formulation as shown below. Let
\begin{equation}
\begin{aligned}
\phi_{1}=& \phi_{D}-\phi_{L}, \\ 
\phi_{2} =&
\begin{cases} 
      -\left( \phi_{R}-\phi_{U}\right)  & \text{for bosons} \\
      -\left( \phi_{R}-\phi_{U}\right)  + \frac{\pi}{2} & \text{for fermions} .
   \end{cases}
\end{aligned}
\end{equation}
The generalized detection probabilities of Eqs.~\eqref{path_path},~\eqref{spin_spin},~\eqref{spin_path}, and~\eqref{path_spin} are, respectively, given by 
\begin{eqnarray}
 \begin{tabular}{c | c c} 
  & $B$ : $R$ & $B$ : $U$  \\  
 \hline
 $A$ : $D$ & \hspace{0.1cm} $\frac{1}{4}\text{cos}^2(\phi_{1}-\phi_{2})$ \hspace{0.1cm} & $\frac{1}{4}\text{sin}^2(\phi_{1}-\phi_{2})$\\ 
 $A$ : $L$ & \hspace{0.1cm}  $\frac{1}{4}\text{sin}^2(\phi_{1}-\phi_{2})$ \hspace{0.1cm} & $\frac{1}{4}\text{cos}^2(\phi_{1}-\phi_{2})$ \\
\end{tabular},\\
 \begin{tabular}{c | c c} 
  & $B$ : $\downarrow$ & $B$ : $\uparrow$  \\  
 \hline
 $A$ : $\downarrow$ & \hspace{0.1cm} $\frac{1}{4}\text{sin}^2(\phi_{1}-\phi_{2})$ \hspace{0.1cm} & $\frac{1}{4}\text{cos}^2(\phi_{1}-\phi_{2})$\\ 
 $A$ : $\uparrow$ & \hspace{0.1cm} $\frac{1}{4}\text{cos}^2(\phi_{1}-\phi_{2})$ \hspace{0.1cm} & $\frac{1}{4}\text{sin}^2(\phi_{1}-\phi_{2})$ \\
\end{tabular},\\
 \begin{tabular}{c | c c} 
  & $B$ : $R$ & $B$ : $U$  \\  
 \hline
 $A$ : $\downarrow$ &\hspace{0.1cm} $\frac{1}{4}\text{sin}^2(\phi_{1}-\phi_{2})$ \hspace{0.1cm} & $\frac{1}{4}\text{cos}^2(\phi_{1}-\phi_{2})$\\ 
 $A$ : $\uparrow$ & \hspace{0.1cm} $\frac{1}{4}\text{cos}^2(\phi_{1}-\phi_{2})$ \hspace{0.1cm} & $\frac{1}{4}\text{sin}^2(\phi_{1}-\phi_{2})$ \\
\end{tabular},\\
 \begin{tabular}{c | c c} 
  & $B$ : $\downarrow$ & $B$ : $\uparrow$  \\  
 \hline
 $A$ : $D$ & \hspace{0.1cm} $\frac{1}{4}\text{cos}^2(\phi_{1}-\phi_{2})$ \hspace{0.1cm} & $\frac{1}{4}\text{sin}^2(\phi_{1}-\phi_{2})$\\ 
 $A$ : $L$ & \hspace{0.1cm} $\frac{1}{4}\text{sin}^2(\phi_{1}-\phi_{2})$ \hspace{0.1cm} & $\frac{1}{4}\text{cos}^2(\phi_{1}-\phi_{2})$ \\
\end{tabular}.
\end{eqnarray}

Computations, following Eqs.~\eqref{Normal} and~\eqref{CHSH}, lead to the maximum violation of Bell's inequality. 

\begin{figure}[t!] 
\centering
\includegraphics[width=7.6cm]{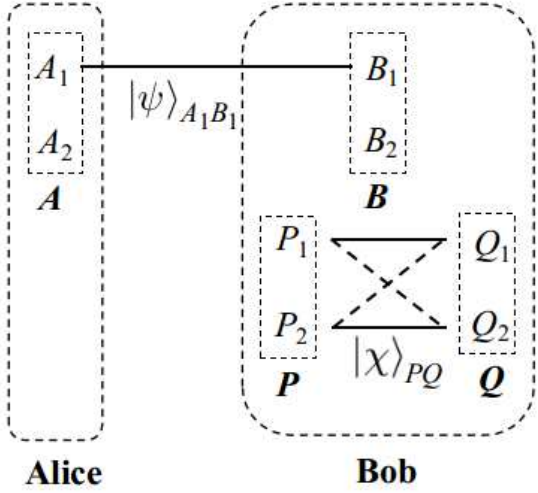} 
\caption{The singlet state $\ket{\psi}_{A_{1}B_{1}}$ is shared between DOF 1 of Alice and that of Bob, whereas HHES $\ket{\chi}_{PQ}$ is kept by Bob.}
\label{fig:Alice_Bob}
\end{figure}

\section{Does the scheme of Li \textit{et al.}~\cite{HHNL} work for distinguishable particles?} \label{DisHHES}
We are interested to see whether the circuit of Li \textit{et al.}~\cite[Fig. 2]{HHNL} gives the same results for two distinguishable particles. Let us calculate the term in the first row and first column of Eq.~\eqref{path_path} for fermions. It says that the probability of Alice detecting a particle in detector $D$ and Bob detecting a particle in detector $R$ is given by
\begin{equation}
\left| \frac{1}{4} \left[  e^{i\left( \phi_{R}+ \phi_{L} \right) } + e^{i\left( \phi_{D}+ \phi_{U} \right) } \right] \right|^{2}=\frac{1}{4} \text{cos}^{2}\phi.
\end{equation}
If the particles are made distinguishable, this probability is calculated as
\begin{equation}
 \left| \frac{1}{4} e^{i\left( \phi_{R}+ \phi_{L} \right) } \right|^{2} + \left| \frac{1}{4} e^{i\left( \phi_{D}+ \phi_{U} \right) } \right|^{2} =\frac{1}{8}.
\end{equation}
As for other terms of Eq.~\eqref{path_path}, each term of Eqs~\eqref{spin_spin},~\eqref{spin_path}, and~\eqref{path_spin} reduces to $\frac{1}{8}$. From that, one can easily show that the right hand side of Eq.~\eqref{Normal} becomes zero. Thus the Bell violation is not possible by the CHSH test. Similar calculations for the bosons lead to the same conclusion. So, the circuit of~\cite{HHNL} would not work for distinguishable particles.

\section{No HHES for distinguishable particles} \label{NoHHES}
 It is well-known that UFQT~\cite{Popescu94} for distinguishable particles is possible using BSM and LOCC. Here, we show that if HHES for distinguishable particles could exist, then one could construct a universal quantum cloning machine (UQCM)~\cite{QC05,QC14} using UFQT and HHES, and further, use that UQCM to achieve signaling.

\subsection{Our signaling protocol} Our signaling protocol works in three phases, as follows.
\begin{figure}[t!] 
\centering
\includegraphics[width=7.6cm]{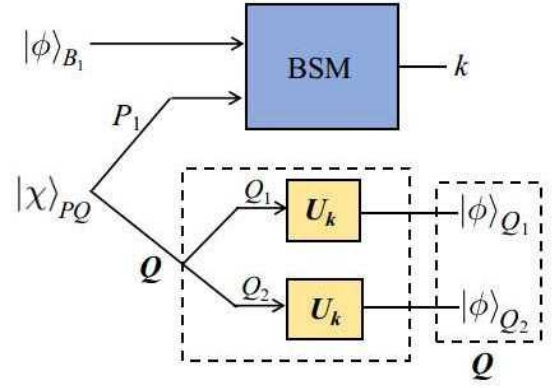} 
\caption{Our proposed universal quantum cloning machine to demonstrate the impossibility of HHES using distinguishable particles. Inputs to this cloning machine are the unknown quantum state $\ket{\phi}$ of DOF 1 of the particle $B$ (denoted by $\ket{\phi}_{B_{1}}$) and the hyper-hybrid entangled state $\ket{\chi}_{PQ}$, as shown in Fig.~\ref{fig:Alice_Bob}. More specifically, $\ket{\phi}_{B_{1}}$ is the state unknown to Bob generated on Bob's side after Alice does measurement on $A_1$ in either $Z$ basis or $X$ basis. After that, Bob performs BSM on $B_1$ and $P_1$, resulting in one of the four possible Bell states as output, denoted by $k$. Based on this output $k$, suitable unitary operations $U_{k}$ are applied on both the DOFs of $Q$, i.e., $Q_1$ and $Q_2$, where $U_{k} \in \left\lbrace \mathcal{I}, \sigma_{x}, \sigma_{y}, \sigma_{z}\right\rbrace $, $\mathcal{I}$ being the identity operation and $\sigma_{i}$'s $\left( i = x, y, z \right)$ the Pauli matrices. As a result, the unknown state $\ket{\phi}$ of $B_1$ is copied to both $Q_1$ and $Q_2$.}
\label{fig:NO_CLONING}
\end{figure}

\subsubsection{ First phase: Initial set-up.} 
Suppose there are four particles $A$, $B$, $P$, and $Q$ each having two DOFs $1$ and $2$. The particle $A$ is with Alice and the remaining three are with Bob, who is spacelike separated from Alice. The pair $\left\lbrace A,B \right\rbrace$ is in the singlet state in DOF $1$  denoted by $\ket{\psi}_{A_{1}B_{1}}$ and the pair $\left\lbrace P,Q \right\rbrace$ is in HHES using both the DOFs $1$ and $2$ denoted by $\ket{\chi}_{PQ}$. To access the DOF $i$ of particle $X$, we use the notation $X_i$, where $X \in \{A, B, P, Q\}$ and $i \in \{1, 2\}$. The situation is depicted in Fig.~\ref{fig:Alice_Bob}. 
Note that $\ket{\psi}_{A_{1}B_{1}}$ can be expressed in any orthogonal basis. We have taken only $Z$ basis or computational basis $\left\lbrace \ket{0}, \ket{1}\right\rbrace$ and $X$ basis or Hadamard basis $\left\lbrace \ket{+}, \ket{-}\right\rbrace$ such that 
\medmuskip=2mu
\thinmuskip=2mu
\thickmuskip=2mu 
\begin{equation}
\begin{aligned}
   \ket{\psi}_{A_{1}B_{1}}&= \frac{1}{\sqrt{2}} \left( \ket{01}_{A_{1}B_{1}}-\ket{10}_{A_{1}B_{1}}\right)\\ & =  \frac{1}{\sqrt{2}} \left(\ket{+-}_{A_{1}B_{1}}-\ket{-+}_{A_{1}B_{1}}\right). 
\end{aligned}
\end{equation}
\medmuskip=2mu
\thinmuskip=2mu
\thickmuskip=2mu 
Alice wants to transfer binary information instantaneously to Bob. Before going apart, Alice and Bob agree on the following convention.
\begin{enumerate}
\item If Alice wants to send zero to Bob, then she would measure in $Z$ basis on the DOF 1 of her particle so that the state of the DOF 1 of the particle at Bob's side would be either $\ket{0}$ or $\ket{1}$. 
\item If Alice wants to send $1$ to Bob, then she would measure in $X$ basis on the DOF 1 of her particle so that the state of the DOF 1 of the particle at Bob's side would be either $\ket{+}$ or $\ket{-}$.
\end{enumerate}

\subsubsection{Second phase: Cloning of any unknown state.} 
There are two steps of our proposed UQCM as follows.
\begin{enumerate}
\item Alice does measurement on DOF 1 of her particle $A$, i.e., $A_1$ in either $Z$ basis or $X$ basis. After this measurement, the state on DOF 1 of particle $B$, i.e., $B_1$ on Bob's side, is in an unknown state $\ket{\phi} \in \left\lbrace \ket{0}, \ket{1}, \ket{+}, \ket{-} \right\rbrace$ and it is  denoted  by $\ket{\phi}_{B_1}$. 
\item After Alice's measurement, Bob performs BSM on DOF 1 of particles $B$ and $P$,  i.e., on $B_1$ and $P_1$. 
This results in an output $k$ as one of the four possible Bell states (as seen in standard teleportation protocol~\cite{QT93}). Based on this output $k$, suitable unitary operations $U_{k}$ are applied on both the DOFs of $Q$, i.e., $Q_1$ and $Q_2$, where $U_{k} \in \left\lbrace \mathcal{I}, \sigma_{x}, \sigma_{y}, \sigma_{z}\right\rbrace $, $\mathcal{I}$ being the identity operation and $\sigma_{i}$'s $\left( i = x, y, z \right)$ the Pauli matrices.
As the first DOF of particle $P$, $P_1$ is maximally entangled with both the DOFs of $Q$, i.e., $Q_1$ and $Q_2$; thus, using BSM on DOFs $B_1$ and $P_1$ and suitable unitary operations on DOFs $Q_1$ and $Q_2$, the unknown state $\ket{\phi}$ on DOF ${B_1}$ is copied to both the DOFs $Q_1$ and $Q_2$. This part of the circuit, shown in Fig.~\ref{fig:NO_CLONING}, acts as a UQCM. 
\end{enumerate}

\subsubsection{Third phase: Decoding Alice's measurement basis.}
Now from the two copies of the unknown state $\ket{\phi}$ on the two DOFs of $Q$,  i.e., $\ket{\phi}_{Q_1}$ and $\ket{\phi}_{Q_2}$, Bob tries to discriminate the measurement bases of Alice, so that he can decode the information sent to him. For that, Bob measures both the DOFs of $Q$ in $Z$ basis, resulting in either $0$ or  $1$ in each of the DOFs. Now there are two possibilities.
\begin{enumerate}
\item If Alice has measured in $Z$ basis, then Bob's possible measurement results on the two DOFs of $Q$ are $\{00, 11\}$. 
\item On the other hand, if Alice has measured in $X$ basis, then Bob's possible measurement results on the two DOFs of $Q$ are $\{00, 01, 10, 11\}$.
\end{enumerate}
Suppose, Bob adopts the following strategy. Whenever his measurement results are all zero or all one (i.e., 00 or 11), then he concludes that Alice has sent a zero, and whenever he measures otherwise (i.e., 01 or 10) then he concludes that Alice has sent a 1.

\subsection{Computation of the signaling probability}
Let the random variables $X_A$ and $X_B$ denote the bit sent by Alice and the bit decoded by Bob, respectively. Hence, under the above strategy, Bob's success probability of decoding, which is also the probability of signaling, is given by
\begin{equation}
\begin{aligned}
P_{sig}  = & \Pr(X_A = 0 \wedge X_B = 0) + \Pr(X_A = 1 \wedge X_B = 1)\\
 = & \Pr(X_A = 0)\cdot\Pr(X_B = 0 \mid X_A = 0) \\
& + \Pr(X_A = 1)\cdot\Pr(X_B = 1 \mid X_A = 1)\\
 = & \frac{1}{2}\cdot 1 + \frac{1}{2}\cdot\frac{2}{4} = 0.75.
\end{aligned}
\end{equation}

To increase $P_{sig}$ further, Bob can use HHES involving $N$ DOFs of $P$ and $Q$, with $N \geq 3$. Then he can make $N$ copies of the unknown state $\ket{\phi}$ into the DOFs of $Q$. Analogous, to the strategy above for the case $N=2$, here also if all the measurement results of Bob in the $N$ DOFs of $Q$ in $Z$ basis are the same, i.e., all-zero case or the all-one case, then Bob concludes that Alice has sent a zero; otherwise, he concludes that Alice has sent a 1. Thus, the above expression of $P_{sig}$ changes to
$$\frac{1}{2}\cdot 1 + \frac{1}{2}\cdot\frac{2^N-2}{2^N}.$$
In other words, 
\begin{equation}
\label{sigprob}
P_{sig}=1 - \frac{1}{2^N}.
\end{equation}   
By making $N$ larger and larger,  $P_{sig}$ can be made arbitrarily close to 1.

\subsection{Experimental realization of the signaling scheme using HHES of distinguishable particles}
 For the experimental realization of the above protocol, we propose a circuit with DOF sorters, such as a	 spin sorter (SS), path sorter (PS), etc. (A spin sorter can be realized in an optical system using a polarizing beam splitter for sorting between the horizontal $\ket{H}$ and the vertical $\ket{V}$ polarizations of a photon. For an
alternative implementation in atomic systems using Raman process, one can see~\cite[Fig.~3]{HHNL}.)

\begin{figure}[t!] 
\centering
\includegraphics[width=8.6cm]{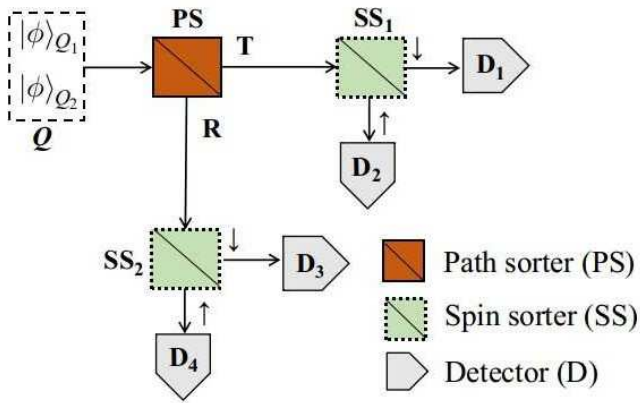} 
\caption{Bob's circuit for distinguishing between $Z$ and $X$ bases using two types of DOF sorters, i.e., a spin sorter (SS) and a path sorter (PS), and four detectors. Here the output of Fig.~\ref{fig:NO_CLONING} is used as an input in this circuit. Here, $\ket{\downarrow}$ and $\ket{\uparrow}$ denote the down- and the up-spin states of the particles, respectively. Further, $\ket{T}$ denotes the transverse mode and $\ket{R}$ denotes the reflected modes of a PS.}
\label{SLS_N_2}
\end{figure}

Suppose DOFs $1$ and $2$  are spin and path, respectively, with the two output states $\left\lbrace \ket{\downarrow}, \ket{\uparrow} \right\rbrace $ and $\left\lbrace \ket{T},\ket{R}\right\rbrace $. Here, $\ket{\downarrow}$ and $\ket{\uparrow}$ denote the down and the up spin states of the particles and $\ket{T}$ and $\ket{R}$ denote the transverse and the reflected modes of a PS, respectively. Without loss of generality, we take 
\begin{equation}
\begin{aligned}
\ket{0}=&\ket{\downarrow}=\ket{T}, \hspace{0.2cm} \ket{1}=\ket{\uparrow}=\ket{R}, \\  \ket{+}=&\frac{1}{\sqrt{2}}\left( \ket{\downarrow}+\ket{\uparrow}\right) =\frac{1}{\sqrt{2}}\left(\ket{T}+\ket{R}\right), \\
\ket{-}=&\frac{1}{\sqrt{2}}\left(\ket{\downarrow}-\ket{\uparrow}\right) =\frac{1}{\sqrt{2}}\left(\ket{T}-\ket{R}\right).
\end{aligned}
\end{equation}
The circuit, shown in Fig.~\ref{SLS_N_2}, takes as input particle $Q$ with DOFs $1$ and $2$, each having the cloned state $\ket{\phi}$ from the output of the circuit in Fig.~\ref{fig:NO_CLONING}. Bob places a path sorter $PS$ followed by two spin sorters $SS_{1}$ and $SS_{2}$ on two output modes of $PS$. Let $ D_{1}$ ($ D_{3}$) and $D_{2}$ ($ D_{4}$) be the detectors at the two output ports of $SS_{1}$ ($SS_{2}$).

If Alice measures in $Z$ basis, Bob detects the particles in $\left\lbrace D_{1},D_{4}\right\rbrace$ with unit probability. On the other hand, if she measures in $X$ basis, the particles would be detected in each of the detector sets $\left\lbrace D_{1},D_{4}\right\rbrace$ and $\left\lbrace D_{2},D_{3}\right\rbrace $ with a probability of 0.5. When Bob detects the particles in either $D_2$ or $D_3$, he instantaneously knows that the measurement basis of Alice is $X$. In this case, the signaling probability is 0.75, which can be obtained by putting $N=2$ in Eq.~\eqref{sigprob}.

\begin{figure}[t!] 
\centering
\includegraphics[width=8.6cm]{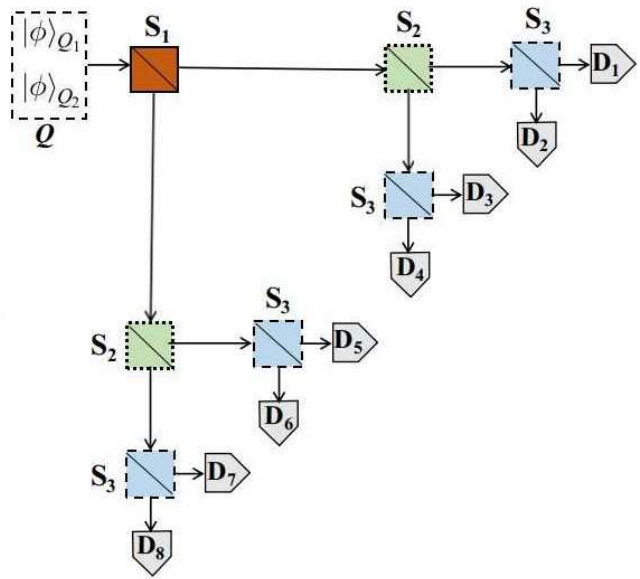} 
\caption{Bob's circuit for distinguishing between $Z$ and $X$ bases using three types of DOF sorters, i.e., $S_1$, $S_2$, and $S_3$, and eight detectors. Here the output of Fig.~\ref{fig:NO_CLONING} is used as an input in this circuit.}
\label{SLS_N_3}
\end{figure}

For better signaling probability, one can use three DOFs $1$, $2$, and $3$, instead of two, in the joint state $\ket{\chi}_{PQ}$. The scheme for three DOFs is shown in Fig.~\ref{SLS_N_3}, where $S_{i}$ represents the sorter for DOF $i$, for $i \in \{1, 2, 3\}$. Now, if Alice  measures in $Z$ basis, Bob detects the particles in $\left\lbrace D_{1},D_{8}\right\rbrace $ with probability 1. But if she measures in $X$ basis, the particles would be detected in $\left\lbrace D_{1},D_{8}\right\rbrace $ with probability $\frac{2}{2^3}$ and in  $\left\lbrace D_{2}, \cdots , D_{7}\right\rbrace $ with probability $\frac{2^3 -2}{2^3}=0.75$. In this case, the signaling probability is 0.875, which can be obtained by putting $N=3$ in Eq.~\eqref{sigprob}.

We can generalize the above schematic as follows. Suppose, each of $P$ and $Q$ has $N$ DOFs (each degree having two eigenstates), numbered 1 to $N$, in HHES $\ket{\chi}_{PQ}$. We also need to use $N$ corresponding types of DOF sorters. Now, if Alice measures in $Z$ basis, Bob detects the particles in $\left\lbrace D_{1},D_{2^N}\right\rbrace $ with probability 1. But if she measures in $X$ basis, the particles are detected in $\left\lbrace D_{1},D_{2^N}\right\rbrace $ with probability $\frac{2}{2^N}$ and in detectors  $\left\lbrace D_{2}, \cdots , D_{2^N-1}\right\rbrace $ with probability $\frac{2^N-2}{2^N}$. For $N$ DOFs, the signaling probability is given in Eq.~\eqref{sigprob}.

\subsection{Increasing signaling probability without increasing the number of DOFs}
From Eq.~\eqref{sigprob}, it is clear that the signaling is possible only when $N$ is infinitely large. The existence of such a huge number of accessible DOFs may be questionable. Interestingly, we devise an alternative schematic that can drive the asymptotic success probability to 1 with only two DOFs but using many copies of the singlet state shared between Alice and Bob and the same number of copies of HHES at Bob's disposal.

Suppose, Alice and Bob share $M$ copies of the singlet state $\left\lbrace  \ket{\psi^{(1)}}_{A_{1}B_{1}}, \ket{\psi^{(2)}}_{A_{1}B_{1}}, \cdots, \ket{\psi^{(M)}}_{A_{1}B_{1}}  \right\rbrace $ and Bob also has an equal number of copies of HHES $ \left\lbrace \ket{\chi^{(1)}}_{PQ}, \ket{\chi^{(2)}}_{PQ}, \cdots, \ket{\chi^{(M)}}_{PQ}  \right\rbrace $ (a single copy of the singlet state and HHES is shown in Fig.~\ref{fig:Alice_Bob}). Now the cloning can be performed in the following two steps.
\begin{enumerate}
\item  Alice performs measurement in her preferred basis on each DOF 1 of her $M$ particles so that on the DOF 1 of each of the $M$ particles on Bob's side, a copy of the unknown state $\ket{\phi}$ is obtained.
\item After that, Bob performs BSM on each of the $M$ pairs of $\left\lbrace B_1, P_1 \right\rbrace $ and suitable unitary operations so that the unknown state $\ket{\phi}$ is copied to each of the $M$ pairs of $\left\lbrace Q_1, Q_2 \right\rbrace $.
\end{enumerate}

Now Bob passes each of the $M$ copies of $Q$ as shown in Fig.~\ref{SLS_N_2} and adopts the following strategy. If Bob receives each of the $M$ particles in $D_{1}$ or $D_{4}$, then he concludes that Alice has measured in $Z$ basis. On the other hand, if Bob observes any one of the $M$ particles in $D_{2}$ or $D_{3}$, then he concludes that Alice has measured in $X$ basis. Under this strategy, Bob encounters a decoding error whenever Alice has measured in $X$ basis, but he receives all the $M$ particles in $D_{1}$ or $D_{4}$. 
In this case, the probability that a single particle is detected in the detector set $\left\lbrace D_{1},D_{4}\right\rbrace$ is $\frac{1}{2}$ and hence the probability that all the $M$ particles are detected in the above set is $\frac{1}{2^M}$. Hence, the corresponding success probability of signaling is given by 
\begin{equation}
    P_{sig}=1-\frac{1}{2^M},
\end{equation}
which also asymptotically goes to 1.

The essence of the above discussion is that, in a world where special relativity holds barring signaling, HHES for distinguishable particles is not possible.

\section{No unit fidelity quantum teleportation for indistinguishable particles} \label{NoUFQT}
 Earlier, we have shown that signaling for distinguishable particles can be achieved using UFQT and HHES as black-boxes. UFQT for distinguishable particles is already known~\cite{QT93}, and so we have concluded that HHES for distinguishable particles must be an impossibility.

Using massive identical particles, Marzolino and Buchleitner~\cite{Ugo15} have shown that UFQT is not possible using a finite and fixed number of indistinguishable particles, due to the particle number conservation superselection rule (SSR)~\cite{Wick52,SSR07}. Interestingly, several independent works~\cite{SSR07,Suskind67,Paterek11} have already established that this SSR can be bypassed. So, an obvious question is: whether it is possible to perform UFQT for indistinguishable particles bypassing the SSR. This question is also answered in the negative in~\cite{Ugo15}. 

Very recently, for indistinguishable particles, Lo Franco and Compagno~\cite{LFC18} have achieved a quantum teleportation fidelity of $5/6$, overcoming the classical teleportation fidelity bound $2/3$~\cite{HHH99}. But they have not proved whether this value is optimal or whether unit fidelity can be achieved or not.

Systematic calculations show that our earlier scheme of quantum cloning and signaling (Figs.~\ref{fig:Alice_Bob} and Fig.~\ref{fig:NO_CLONING}) would still work, even if one replaces the UFQT and HHES tools for distinguishable particles with those of the indistinguishable ones (assuming that such tools exist). As signaling is not possible in quantum theory, we have concluded that HHES for distinguishable particles is not possible. But, for indistinguishable particles, the creation of HHES is possible~\cite{HHNL}. Thus, a logical conclusion is that, to prevent signaling, UFQT must not be possible for indistinguishable particles.

\section{A separation result between distinguishable and indistinguishable particles} \label{Sepresult}
\begin{figure}[h!]
\centering
\includegraphics[width=8cm]{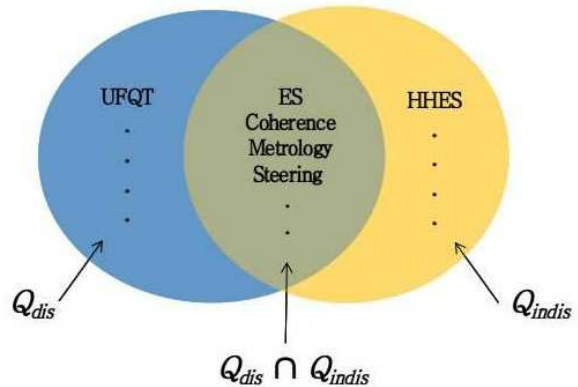} 
\caption{The two sets $Q_{dis}$ (consisting of quantum properties and applications of distinguishable particles), $Q_{indis}$ (consisting of quantum properties and applications of indistinguishable particles), and their intersection (UFQT, HHES, and ES stand for unit fidelity quantum teleportation, hyper-hybrid entangled state, and entanglement swapping, respectively).} 
\label{fig:Overall}
\end{figure}

From the above two no-go results, we can establish a separation result between distinguishable and indistinguishable particles.  
Let $Q_{dis}$ and $Q_{indis}$ be the two sets consisting of quantum properties and applications of distinguishable and indistinguishable particles, respectively, as shown in Fig.~\ref{fig:Overall}.

Several earlier works have attempted extending many results on one of these two sets to the other. For example, quantum teleportation was originally proposed for distinguishable particles~\cite{QT93,QTnat15}. But recent works~\cite{LFC18,Ugo15} have extended it for indistinguishable particles. Similarly, duality of entanglement as proposed in~\cite{Bose13} was thought to be a unique property of indistinguishable particles. But later its existence for distinguishable particle was shown in~\cite{M.Karczewski16}. Another unique property of quantum correlation is quantum coherence, which was proposed for distinguishable particles in~\cite{Baumgratz14} and later for indistinguishable particles in~\cite{Sperling17}. Einstein-Podolsky-Rosen steering~\cite{Schrodinger} was extended from distinguishable particles to a special class of indistinguishable particles called Bose-Einstein condensates~\cite{Fadel18}. Entanglement swapping, originally proposed for distinguishable particles in~\cite{ES93,Pan10,Pan19}, was also shown for indistinguishable particles in~\cite{LFCES19}.

To the best of our knowledge, there is no known quantum correlation or application that is unique for distinguishable particles only and does not hold for indistinguishable particles, and vice versa. In Sec.~\ref{NoHHES} of this paper, we have shown that HHES is unique to the set $Q_{indis}$, and in Sec.~\ref{NoUFQT} we have established that UFQT is unique to the set $Q_{dis}$. Thus, we demonstrate a clear separation between these two sets.

\section{Entanglement Swapping using only two indistinguishable particles without BSM} \label{2PES}
 Here we present an ES protocol using two indistinguishable particles, say, $A$ and $B$, without BSM by suitably modifying the circuit of Li \textit{et al.}~\cite{HHNL}. The basic idea is to use any method that destroys the identity of the individual particles, like particle exchange~\cite{Y&S92PRA,Y&S92PRL} or measurement induced entanglement~\cite{Chou05}. Such methods added with suitable unitary operations transfer the intraparticle hybrid entanglement in $A$ (or $B$) to inter-particle hybrid-entanglement between $A$ and $B$. 

\begin{figure}[h!]
\centering
\includegraphics[width=8.6cm]{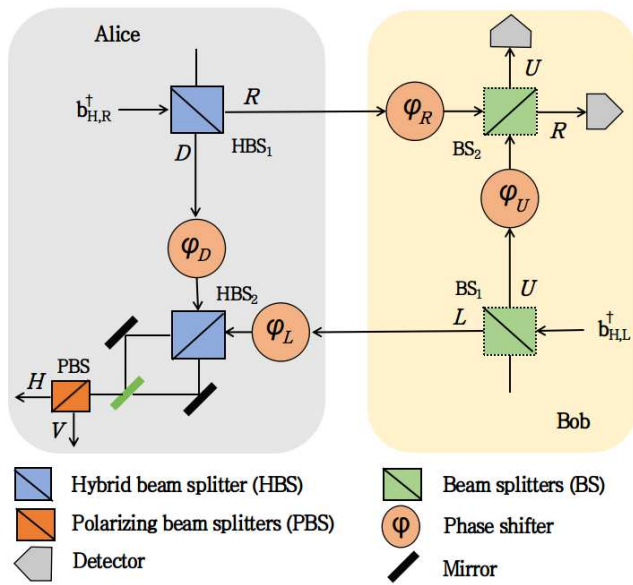} 
\caption{Entanglement swapping with only two indistinguishable particles without Bell state measurements.}
\label{fig:ES}
\end{figure}

Next, we present an optical realization using particle exchange method. Suppose Alice and Bob have two horizontally polarized photons $A$ and $B$, entering into the two modes $R$ and $L$ of a HBS~\cite{HHNL} and a BS, respectively. In second quantization notation, the initial joint state is given by $\ket{\Psi_{i}}=b^{\dagger}_{H,R}b^{\dagger}_{H,L}\ket{0}$, where $\ket{H}$ and $\ket{V}$ denote horizontal and vertical polarization, respectively, and $b_{H,R}$ and $b_{H,L}$ are the corresponding bosonic creation operators satisfying the canonical commutation relations:
\begin{equation}
\left[  b_{i,\textbf{p}_{i}}, b_{j,\textbf{p}_{j}} \right]  = 0,  \left[  b_{i,\textbf{p}_{i}}, b^{\dagger}_{j,\textbf{p}_{j}}\right]  = \delta(\textbf{p}_{i}-\textbf{p}_{j})\delta_{ij}.
\end{equation}

 After passing through HBS$_{1}$, Alice's photon is converted into intraparticle hybrid-entangled state $\frac{1}{\sqrt{2}}\left( b^{\dagger}_{H,R} + i b^{\dagger}_{V,D}\right) $. The particle exchange operation is performed between Alice and Bob, such that the photons coming from $D$ ($U$) and $L$ ($R$) mode go into Alice's (Bob's) side. Next, Alice applies path dependent (or polarization dependent) phase shifts $\varphi_{D}$ and $\varphi_{L}$ on the photons coming from her and Bob's parts, respectively, which go into  HBS$_2$. Similarly, Bob applies path-dependent phase shifts $\varphi_{U}$ and $\varphi_{R}$ on the photons coming from his and Alice's parts, respectively, which go into BS$_2$ as shown in Fig.~\ref{fig:ES}.
The final state is given by
\begin{equation} \label{ESstate}
\begin{aligned}
&\ket{\Psi_{f}}=\frac{1}{4} \left[ e^{i\varphi_{R}}\left( b^{\dagger}_{H,R}+ib^{\dagger}_{H,U}\right) +ie^{i\varphi_{D}} \left( b^{\dagger}_{V,D} + i b^{\dagger}_{H,L}\right)  \right] \\
& \otimes \left[ e^{i\varphi_{L}} \left( b^{\dagger}_{H,L}+ib^{\dagger}_{V,D}\right)  + i e^{i\varphi_{U}}\left( b^{\dagger}_{H,U}+ib^{\dagger}_{H,R}\right) \right] \ket{0}. 
\end{aligned}
\end{equation}
If Alice measures in the polarization DOF and Bob measures in the path DOF, from Eq.~\eqref{ESstate} the probabilities that both of them detect one particle are given by  
\begin{equation}
 \begin{tabular}{c | c c} 
  & B : R & B : U  \\  
 \hline
 A : H \hspace{0.1cm} & $ \frac{1}{4} \text{cos}^{2} \varphi $ \hspace{0.1cm} & $\frac{1}{4} \text{sin}^{2} \varphi$ \\
 A : V \hspace{0.1cm} & $\frac{1}{4} \text{sin}^{2} \varphi$ \hspace{0.1cm} &  $\frac{1}{4} \text{cos}^{2} \varphi$ \\
\end{tabular},
\label{Table:1}
\end{equation}
where $\varphi=\left( \varphi_{D} -\varphi_{L} - \varphi_{R}+ \varphi_{U}\right) / 2$. With suitable values of the phase shifts, one can get the Bell violation in the CHSH~\cite{CHSH} test up to Tsirelson's bound~\cite{Tsirelson}. It can be easily verified that after particle exchange, the particle received at Alice's side (or Bob's side) has no hybrid entanglement, because it is transferred between them. 

Note that the difference between the circuit of Li \textit{et al.}~\cite[Fig. 2]{HHNL} and  Fig.~\ref{fig:ES} is that both the HBSs in Bob's side in the former circuit are replaced by BSs in the latter.
Thus, the intraparticle hybrid entanglement of one particle is transferred into the inter-particle hybrid-entanglement of two particles.

\section{Conclusion} \label{Conclusion}
In this paper, we settle several important open questions that arise due to the recent work on hyper-hybrid entanglement for two indistinguishable bosons by Li \textit{et al.}~\cite{HHNL}.

In particular, we have shown that such entanglement can also exist for two indistinguishable fermions. Further, we have argued that, if in their circuit the particles are made distinguishable, such type of entanglement vanishes. We have also proved the following two no-go results---no HHES for distinguishable particles and no UFQT for indistinguishable particles---as in either case the no-signaling principle is violated. 

Finally, we have shown an application where indistinguishable particles provide more efficiency than distinguishable ones in terms of resource, contrary to the existing results. This application an entanglement swapping  protocol with only  two  indistinguishable  particles  using linear optics where Bell state measurement is not required. 

Our results establish that there exists some quantum correlation or application unique to indistinguishable particles only and yet some unique to distinguishable particles only, giving
a separation between the two domains.

The present results can motivate researchers to find more quantum correlations and applications that are either unique to distinguishable or indistinguishable particles or applicable to both. For the latter case, a comparative analysis of the resource requirements and the efficiency or fidelity can also be a potential future work.

\section*{ACKNOWLEDGMENT}
Anindya Banerji acknowledges R. C. Bose Centre for Cryptology and Security of Indian Statistical Institute Kolkata for hosting his visit during February-March of 2019, when part of this research work was done.

\appendix*

\section{DEFINITION OF ENTANGLEMENT}
\label{app}
Here, we discuss the definition of entanglement of distinguishable and indistinguishable particles as proposed in~\cite{Benatti10,Benatti12,Benatti14} and followed in our current paper.

Let us consider a many-body system which is represented by the Hilbert space $\mathcal{H}$. The algebra of all bounded operators, which includes all the observables, is represented by $\mathcal{Z(H)}$. In this algebraic framework, the standard notions of states and the tensor product partitioning of $\mathcal{H}$ are changed into the observables and local structures of $\mathcal{Z(H)}$. Now before defining entanglement, we define \textit{algebraic bipartition} and \textit{local operators}. 

\noindent{\bf Algebraic bipartition}: 
An algebraic bipartition of operator algebra $\mathcal{Z(H)}$ is any pair ($\mathcal{A}$, $\mathcal{B}$) of commuting subalgebras of $\mathcal{Z(H)}$ such that $\mathcal{A}$, $\mathcal{B}\in \mathcal{Z(H)}$. If any element of $\mathcal{A}$ commutes with any element of $\mathcal{B}$, then $\left[  \mathcal{A} ,\mathcal{B} \right]  = 0$.

\noindent{\bf Local operators}:
 For any algebraic bipartition $(\mathcal{A}, \mathcal{B})$, an operator is called a local operator if it can be represented as the product $A B$, where $A \in \mathcal{A}$ and $B \in \mathcal{B}$. 

\noindent{\bf Entangled states}:
For any algebraic bipartition $(\mathcal{A}, \mathcal{B})$, a state $\rho$ on the algebra $\mathcal{Z(H)}$ is called separable if the
expectation of any local operator $A B$ can be decomposed
into a linear convex combination of products of local expectations, as follows
\begin{equation}
\begin{aligned}
 Tr(\rho A B) &= \sum_{k} \lambda_{k}Tr(\rho_{k}^{(1)}A) Tr(\rho_{k}^{(2)} B), \\
  \lambda_{k} & \geq  0,  \hspace{1cm} \sum_{k} \lambda_{k}=1,
\end{aligned}
\end{equation}
where $\rho_{k}^{(1)}$ and $\rho_{k}^{(2)}$ are given states on $\mathcal{Z(H)}$; otherwise the
state $\rho$ is said to be entangled with respect to the algebraic bipartition $(\mathcal{A}, \mathcal{B})$.
Note that, this algebraic bipartition can also be spatial modes like distinct laboratories each controlled by Alice and Bob. If any state cannot be written in the above form, then it would certainly violate the CHSH inequality. Therefore, we use the violation of the CHSH inequality as an indicator of entanglement.

\end{document}